\documentclass[journal,12pt,onecolumn,draftclsnofoot]{IEEEtran}
\usepackage{amsmath,cite,bm,amsfonts,amssymb,graphicx,color,mathtools,
setspace,comment,multirow,xfrac,lipsum,bbm}
\usepackage[justification=centering]{caption}
\usepackage{algorithmic}
\usepackage{algorithm}
\usepackage{citesort}
\usepackage{caption}
\usepackage{setspace}
\usepackage{subcaption}
\usepackage{epstopdf}
\usepackage[T1]{fontenc}

\DeclarePairedDelimiter\floor{\lfloor}{\rfloor}
\DeclarePairedDelimiter{\ceil}{\lceil}{\rceil}
\newcommand*{\QED}{\hfill\ensuremath{\blacksquare}}
\newcommand*\varhrulefill[1][0.4pt]{\leavevmode\leaders\hrule 
height#1\hfill\kern0pt}
\ifCLASSINFOpdf
\else
\fi
\begin{document}

\title{RF Interference in Lens-Based Massive MIMO Systems --- An Application Note}

\author{Harsh Tataria}
\maketitle

\begin{abstract}
We analyze the uplink radio frequency (RF) interference from a multiplicity of single-antenna user equipments transmitting to a cellular base station (BS) within the same time-frequency resource. The BS is assumed to operate with a lens antenna array, which induces additional focusing gain for the incoming signals. Considering line-of-sight propagation conditions, we characterize the multiuser RF interference properties via approximation of the mainlobe interference as well as the effective interferer probability. The results derived in this application note are foundational to more general multiuser interference analysis across different propagation conditions, which we present in a follow up paper. 
\end{abstract}
\vspace{20pt}
\textbf{Notation.} Boldface upper and lower case symbols are used to denote matrices and vectors, while lightface upper and lower case symbols denote scalar quantities. Note that the Hermitian transpose is denoted by $\left(\cdot\right)^{\textrm{H}}$. 
The scalar norm is denoted as $\left|\cdot\right|$ and the floor, ceiling and indicator functions are expressed as $\floor{\cdot}$, $\ceil{\cdot}$, and $\mathbbm{1}\left(\cdot\right)$, respectively. The ``$\textrm{sinc}$" function is given by $\textrm{sinc}\left(x\right)=\sin\left(x\right)/x$, while the ``maximum" and ``minimum" functions are given by $\max\left(\cdot\right)$ and $\min\left(\cdot\right)$. Finally, $\mathcal{O}\left(\cdot\right)$ denotes the ``order" of a mathematical term.  

\section{System Model}
\label{SystemModel}
We consider the uplink of a single-cell system, where the base station (BS) is equipped with a large lens antenna array containing $M$ elements. Multiuser operation is assumed, where the lens array receives uplink data streams from $L$ single-antenna user terminals within the same time-frequency interval. The user terminals are located with a uniform random distribution in area covering a net sector of $2\pi/3$ radians. The array at the BS consists of a flat electromagnetic (EM) lens with elements that are located on the focal arc of the lens. Without loss of generality, we consider azimuth direction-of-arrivals (DOAs) and assume that the flat lens is employed with negligible thickness.\footnote{For simplicity, the elevation DOAs are assumed to be zero, which is practically valid if the relative height difference between 
the transmitter and the receiver is much smaller than their separation distance.} The EM lens array has a total aperture of $D_{y}\times{}D_{z}$, as the array is located on the $y-z$ plane and is centered at the origin. The focal arc of the lens is defined as a semi-circle around the lens's center in the azimuth plane ($x-y$ plane) with radius $F$. Here $F$ physically represents the focal length of the lens. According to this, each element's locations with respect to the lens's phase center can be written as $B_{m}$ ($x_{m}=F\cos{}\left(\theta_{m}\right), y_{m}=-F\sin\left(\theta_{m}\right), z_{m}=0$), where $\theta_{m}\in{}\left[\frac{-\pi}{2},\frac{\pi}{2}\right]$ is the angle of the $m$-th antenna element 
with respect to the $x$-axis, where $m\in{}\mathcal{M}$. Note that $\mathcal{M}=\left\{0,\pm{}1,\dots{},\pm{}\frac{M-1}{2}\right\}$ denotes the set of antenna indices in the lens array.\footnote{For simplicity, it is worth noting that $M$ is assumed to be an odd integer in this study.} The antenna elements are deployed on the focal arc such that $\tilde{\theta}_{m}=\sin\left(\theta_{m}\right)$ are equally 
spaced in the interval $\left[-1,1\right]$ as indicated in \cite{ZENG1}. Doing this yields 
\begin{equation}
\label{thetamtilde}
\tilde{\theta}_{m}=\frac{m}{\tilde{D}},\hspace{5pt} m\in{}\mathcal{M}, 
\end{equation}
where $\tilde{D}=\frac{D_{y}}{\lambda}$ is the lens dimension along the azimuth plane normalized by the carrier wavelength,  
$\lambda$. According to this formulation, more elements are deployed in the center of the array than those on either sides. The relationship between $M$ and $\tilde{D}$ can be observed from \eqref{thetamtilde} as $M=1+\floor{2\tilde{D}}$. As in \cite{ZENG1}, when the lens array receives an uplink signal in the form of a uniform plane wave from terminal $\ell$, with an azimuth DOA $\phi_{\ell}$, the 
resultant signal received by the $m-$th element of the array can be written as 
\begin{equation}
\label{arrayfactor}
a_{m}\left(\phi_{\ell}\right)\approx{}e^{-j\Phi_{0}}\sqrt{A}\hspace{2pt}\left[\textrm{sinc}\left(m-\tilde{D}\tilde{\phi}_{\ell}\right)\right], 
\hspace{5pt} m\in{}\mathcal{M}, 
\end{equation}
where $A=\frac{D_{y}D_{z}}{\lambda^{2}}$ is the normalized aperture, $\Phi_{0}$ is a common phase shift from the lens's 
aperture to the array, and $\tilde{\phi}_{\ell}=\sin{}\left(\phi_{\ell}\right)\in{}\left[-1,1\right]$ is referred to as the spatial frequency 
corresponding to $\phi_{\ell}$. In line with \cite{ZENG1,ZENG2,ZENG3,Xie1}, we assume that the insertion loss of the lens, as well as its boundary effects are negligible.
The expression in \eqref{arrayfactor} across all $M$ antenna elements yields a $M\times{}1$ propagation channel vector from user $\ell$ to the BS, which we denote as $\bm{h}_{\ell}^{\textrm{LOS}}$.

\section{RF Interference Characteristics at the BS}
\label{LineofSightInterferenceCharacteristics}
Considering line-of-sight (LOS) propagation conditions, the received signal strength remains deterministic under maximum-ratio combining processing, while the only 
uncertainty is contained in the interference power via transmission from undesired users to the BS. We denote the total interference power at the BS by $I^{\textrm{LOS}}$ and analyze its form subsequently. 

\textbf{Theorem 1.} Given $M$ antenna elements at the flat lens array, the interference exerted by the $L-1$ interferers on to the desired user $\ell$ can be expressed as $I^{\textrm{LOS}}=\sum\nolimits_{\ell=1}^{L}I_{\ell}^{\textrm{LOS}}$, 
where $I_{\ell}^{\textrm{LOS}}$ is the LOS interference to terminal $\ell$ given by
\begin{align}
\label{singleuserinterference}
I_{\ell}^{\textrm{LOS}}&\hspace{-2pt}=\hspace{-2pt}\frac{1}{M}\left\{\left|\bm{g}_{\ell}^{\textrm{LOS}}\bm{h}_{k}^{\textrm{LOS}}\right|^{2}\right\}\\
\nonumber
&\hspace{-2pt}=\hspace{-2pt}
\frac{1}{M}\hspace{-1pt}\left\{\left|\sum\limits_{m=0}^{M-1}\hspace{4pt}\frac{\frac{A}{2}\hspace{-2pt}
\left[\cos{}\hspace{-3pt}\left(\hspace{-1pt}m\hspace{-2pt}-\hspace{-1pt}\tilde{D}\Delta_{\ell,k}\right)\hspace{-1pt}-\hspace{-1pt}\cos\left(\hspace{-1pt}m\hspace{-2pt}-\hspace{-2pt}
\tilde{D}\tilde{\Delta}_{\ell,k}\right)\right]}{m\hspace{-1pt}\left(\hspace{-1pt}m\hspace{-1pt}-\hspace{-1pt}\tilde{D}\Delta_{\ell,k}\right)\hspace{-1pt}+\hspace{-1pt}\tilde{D}^{2}\hspace{-1pt}\left(\tilde{\phi}_{\ell}\hspace{1pt}\tilde{\phi_{k}}\right)}\right|^{2}\hspace{-1pt}\right\},
\end{align}
where $\bm{g}_{\ell}$ and $\bm{h}_{k}$ are the maximum-ratio combining vector of user $\ell$ and the propagation channel vector from user $k$ to the BS. Note that $\Delta_{\ell,k}=\tilde{\phi}_{\ell}-\tilde{\phi}_{k}$, and $\tilde{\Delta}_{\ell,k}=\tilde{\phi}_{\ell}-\tilde{\phi}_{k}$. 
\begin{figure}[!t]
\vspace{-15pt}
\centering
\hspace{-10pt}
\includegraphics[width=12cm]{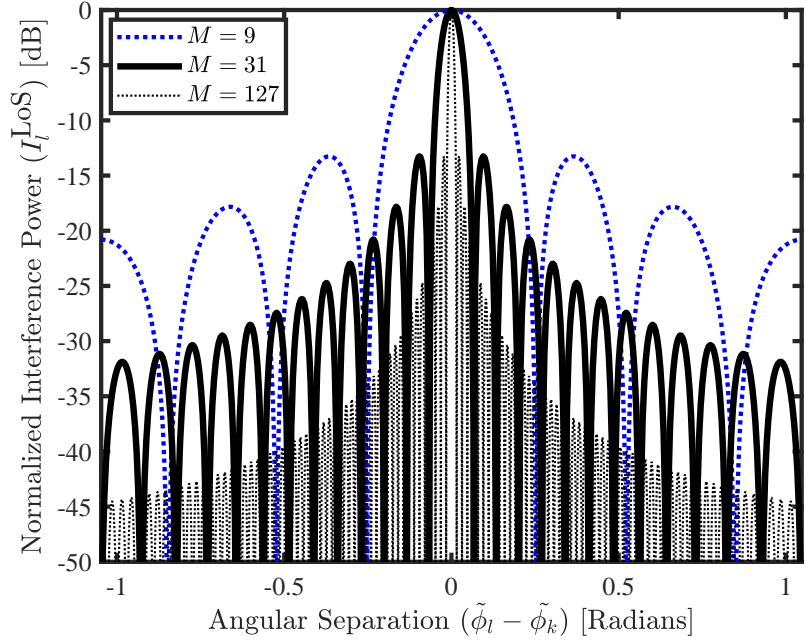}
\caption{LOS Interference power [dB] to terminal $\ell$, $I_{\ell}^{\textrm{LOS}}$ vs. Angular separation of terminals $\ell$ and $k$, $\tilde{\phi}_{\ell}-\tilde{\phi}_{k}$ in [Radians].}
\label{singleuserLoSInterferencePower}
\vspace{-10pt}
\end{figure}

\emph{Proof:} Applying the definition of $\bm{g}_{\ell}^{\textrm{LOS}}$ and $\bm{h}_{k}^{\textrm{LOS}}$, one can recognize that 
$I_{\ell}^{\textrm{LOS}}$ can be written as in equation \eqref{singleuserinterferenceproof1}, shown on the top of the next page for space 
reasons. 
\begin{figure*}[!t]
\begin{align}
\nonumber
I_{\ell}^{\textrm{LOS}}&=\frac{1}{M}\left\{\left|{\left(\bm{h}_{\ell}^{\textrm{LOS}}\right)}^{\textrm{H}}\bm{h}_{k}^{\textrm{LOS}}\right|^{2}\right\}=\frac{1}{M}\left\{\hspace{1pt}\left|
\sum\limits_{m=0}^{M-1}\hspace{-2pt}\sqrt{A}\hspace{2pt}\textrm{sinc}\hspace{-1pt}\left(\hspace{-2pt}m\hspace{-2pt}-\hspace{-2pt}\tilde{D}\sin\hspace{-2pt}\left(\phi_{\ell}\right)\right)\hspace{-3pt}\sqrt{A}\hspace{2pt}\textrm{sinc}\hspace{-2pt}\left(m\hspace{-2pt}-\hspace{-2pt}\tilde{D}\sin\left(\phi_{k}\right)\right)\right|^{2}\right\}\\
\label{singleuserinterferenceproof1}
&\overset{\left(a\right)}=\frac{1}{M}\left\{\hspace{1pt}\left|\hspace{2pt}\sum\limits_{m=0}^{M-1}
\frac{\overbrace{\sqrt{A}\hspace{1pt}\sin\left(\hspace{-2pt}m\hspace{-2pt}-\hspace{-2pt}\tilde{D}\sin\left(\phi_{\ell}\right)\right)
\sqrt{A}\sin\left(m\hspace{-2pt}-\hspace{-2pt}\tilde{D}\sin\left(\phi_{k}\right)\right)}^{=N_{\ell}^{\textrm{LOS}}}}
{\underbrace{\left(m\hspace{-2pt}-\hspace{-2pt}\tilde{D}\sin\left(\phi_{\ell}\right)\right)\hspace{-2pt}
\left(m\hspace{-2pt}-\hspace{-2pt}\tilde{D}\sin\left(\phi_{k}\right)\right)}_{=D_{\ell}^{\textrm{LOS}}}}
\right|^{2}\right\}.
\end{align}
\hrulefill
\end{figure*}
Note that $\left(a\right)$ in \eqref{singleuserinterferenceproof1} is obtained by using the fact that 
$\textrm{sinc}\left(x\right)=\frac{\sin\left(x\right)}{x}$. The numerator and the 
denominator of \eqref{singleuserinterferenceproof1}, in $N_{\ell}^{\textrm{LOS}}$ and $D_{\ell}^{\textrm{LOS}}$, can then be simplified by 
invoking the trigonometric identity $\sin\left(x\right)\sin\left(y\right)=\frac{1}{2}\left[
\cos\left(x\hspace{-2pt}-\hspace{-2pt}y\right)-\cos\left(x\hspace{-2pt}+\hspace{-2pt}y\right)\right]$, where $x=m\hspace{-1pt}-\hspace{-1pt}\tilde{D}\left(\sin\left(\phi_{\ell}\right)\right)$ and 
$y=m\hspace{-1pt}-\hspace{-1pt}\tilde{D}\left(\sin\left(\phi_{k}\right)\right)$, allowing us to write 
\begin{equation}
\label{singleuserinterferenceproof2}
N_{\ell}^{\textrm{LOS}}=\frac{A}{2}\left[\cos{}\left(m-\tilde{D}\left(\tilde{\phi}_{\ell}-\tilde{\phi}_{k}
\right)\right)-\cos\left(m-
\tilde{D}\left(\tilde{\phi}_{\ell}-\left(-\tilde{\phi}_{k}\right)\right)\right)\right], 
\end{equation}
where $\tilde{\phi}_{\ell}=\sin\left(\phi_{\ell}\right)$ and $\tilde{\phi}_{k}=\sin\left(\phi_{k}\right)$, respectively. Using the fact that 
$\Delta_{\ell,k}=\tilde{\phi}_{\ell}-\tilde{\phi}_{k}$ and $\tilde{\Delta}_{\ell,k}=\tilde{\phi}_{\ell}-\left(-\tilde{\phi}_{k}\right)$, 
$N_{\ell}^{\textrm{LOS}}$ in \eqref{singleuserinterferenceproof2} can be re-written as
\begin{equation}
\label{Nlfinal}
N_{\ell}^{\textrm{LOS}}=\frac{A}{2}\left[\cos\left(m-\tilde{D}\Delta_{\ell,k}\right)-\cos\left(m-\tilde{D}\tilde{\Delta}_{\ell,k}\right)\right]. 
\end{equation}
Similarly, after some straightforward algebraic manipulations, $D_{\ell}^{\textrm{LOS}}$ can be expressed as 
\begin{align}
\nonumber
D_{\ell}^{\textrm{LOS}}&=m\hspace{-2pt}\left(\hspace{-1pt}m\hspace{-1pt}-\hspace{-1pt}\tilde{D}\hspace{-1pt}\left(\hspace{-1pt}\tilde{\phi}_{\ell}\hspace{-1pt}-\hspace{-1pt}\tilde{\phi}_{k}\right)\right)\hspace{-1pt}+\hspace{-1pt}\tilde{D}^{2}\hspace{-1pt}\left(\tilde{\phi}_{\ell}\hspace{2pt}\tilde{\phi_{k}}\right)\\
\label{singleuserinterferenceproof3}
&=m\left(m-\tilde{D}\hspace{1pt}\Delta_{\ell,k}\right)+\tilde{D}^{2}\left(\tilde{\phi}_{\ell}\hspace{2pt}\tilde{\phi}_{k}\right). 
\end{align}
Substituting \eqref{singleuserinterferenceproof2} and \eqref{singleuserinterferenceproof3} into \eqref{singleuserinterferenceproof1} yields the desired result and concludes the proof. \QED

\textbf{Remark 1.} By inspecting $I_{\ell}^{\textrm{LOS}}$ in \eqref{singleuserinterference}, one can observe that both the numerator and the denominator contains terms which are expressed as the difference of the $\sin\left(\cdot\right)$ two DOA angles in $\tilde{\phi}_{\ell}$ and $\tilde{\phi}_{k}$. We denote this as the \emph{angular separation} between the DOAs of interferer $k$, and the desired terminal $\ell$. Naturally, when $\tilde{\phi}_{\ell}$ and $\tilde{\phi}_{k}$ are aligned, $I_{\ell}^{\textrm{LOS}}$ will reach its maximum value, inducing maximum spatial correlation in the multiuser channel. In contrast to this, when $\tilde{\phi}_{\ell}$ and $\tilde{\phi}_{k}$ are vastly different, a lower 
value of interference is expected. Furthermore, the RF interference power is also a function of the critical inter-element spacing, $\tilde{D}$ for which the lens array is designed. Rather interestingly, as demonstrated in \cite{TSE1}, the result for $I_{\ell}^{\textrm{LOS}}$ with a half-wavelength spaced uniform linear array yields a very similar conclusion where $N_{\ell}^{\textrm{LOS}}$ and $D_{\ell}^{\textrm{LOS}}$ are both proportional to $\cos(2\pi{}\tilde{D}(\tilde{\phi}_{\ell}-\tilde{\phi}_{k}))$. Note that in that case, the LOS propagation channel has a substantially different form/structure in that it is constructed out of phase shifted exponential functions instead of sinc functions (as for the lens arrays). 

\textbf{Remark 2.} According to \eqref{singleuserinterferenceproof1}, Fig.~\ref{singleuserLoSInterferencePower} depicts $I_{\ell}^{\textrm{LOS}}$ as a function of $\tilde{\phi}_{\ell}-\tilde{\phi}_{k}$. One can observe: 
\begin{enumerate}
\item{$I_{\ell}^{\textrm{LOS}}$ is non-monotonic and non-periodic in nature, making its probability density and cumulative density mathematically intractable to analyze. This is true irrespective of the number of interfering sources present in the system.}
\item{With an increase in the number of elements at the lens array, majority of the interference will lie in the mainlobe (defined from the peak to the first null) of the interference pattern, while the relative sidelobe interference levels are significantly lower. The power ratio between the peak of the main lobe relative to the first sidelobe is approximately 13 dB. For further discussions, see \cite{TATARIA1} and references therein.}
\item{From the result in Theorem 1, the first nulls on either side of the mainlobe appear at $\phi_{\ell}-\phi_{m}=
\pm{}\frac{1}{Md}$. As a result, the mainlobe width can be written as $\frac{2}{Md}$.}
\end{enumerate}

Considering the fact that any further analysis of the instantaneous LOS interference power is intractable, to understand the fundamental nature of its probability and cumulative densities, we approximate it by recognizing that the RF interference, $I_{\ell}^{\textrm{LOS}}$, is composed of a mainlobe surrounded by many smaller sidelobes, and the shape is determined by $M$, the number of antenna elements at the lens array. As the relative sidelobe levels are negligible particularly for moderate and larger arrays, the ``effective" RF interference can be approximated by the mainlobe only. The details of this approximation are as follows, based on which an effective interferer probability is derived. 

\textbf{Approximation 1.} The mainlobe (effective) RF interference can be approximated as 
\begin{equation}
\label{effectiveinterferenceapproximation}
I_{\ell}^{\textrm{LOS}}\approx{}J_{\ell}^{\textrm{LOS}}\hspace{-2pt}.\hspace{3pt}\mathbbm{1}\left(\left|\Theta_{\ell,k}\right|\leq{}1\right),
\end{equation}
where $\Theta_{\ell,k}=\tilde{D}\left(\tilde{\phi}_{\ell}-\tilde{\phi}_{k}\right)$ is the normalized angular separation of terminals $\ell$ and $k$ over half the mainlobe width, such that when $\Theta_{\ell,k}\in{}\left[-1,1\right]$, interferer $k$ falls within the mainlobe of the 
interference pattern. Moreover, the effective interference from interferer $k$, given $M$ elements is 
\begin{align}
J_{\ell}^{\textrm{LOS}}=
\label{jllos}
\frac{1}{M}\left\{\left|\sum\limits_{m=0}^{M-1}\hspace{-2pt}
\frac{\frac{A}{2}\left[\cos\left(m\hspace{-2pt}-\hspace{-2pt}
\Theta_{\ell,k}\right)\hspace{-2pt}-\hspace{-2pt}\cos\left(m\hspace{-1pt}-\hspace{-1pt}\tilde{\Theta}_{\ell,k}\right)\right]}
{m\left(m\hspace{-2pt}-\hspace{-2pt}\Theta_{\ell,k}\right)\hspace{-1pt}+\hspace{-1pt}
\tilde{D}^{2}\left(\tilde{\phi}_{\ell}\hspace{1pt}\tilde{\phi}_{k}\right)}\right|^{2}\right\}, 
\end{align}
with $\tilde{\Theta}_{\ell,k}=\tilde{D}\left(\tilde{\phi}_{\ell}-\frac{\tilde{\phi}_{k}}{-1}\right)$. 

To obtain the density of the interference, we require the probability that an arbitraray interferer (terminal $k$) is an ``effective" interferer, as well as the distribution of the effective interference. Since we are interested in a large (yet naturally a finite element) lens array, we focus on the behavior of the two aforementioned quantities when the number of antenna elements $M$ is large, 
which leads to the following proposition on the effective interferer probability. 

\textbf{Proposition 1.} As $M$ grows, the probability that terminal $k$ is an effective interferer is given by 
\begin{equation}
\label{effectiveinterfererprobability}
p_{k}\approx{}\frac{9}{\tilde{D}\pi^{2}}\left[\tanh^{-1}\left(\frac{\sqrt{3}}{2}\right)\right]. 
\end{equation}

\emph{Proof:} The probability of terminal $k$ being an effective interferer can be written as 
\begin{equation}
\label{effectiveinterfererprob}
p_{k}\approx\mathbb{P}\left(-1\leq{}\Theta_{\ell,k}\leq{}1\right). 
\end{equation}
We recall that all terminals are placed uniformly randomly in a single sector of $\frac{2\pi}{3}$ radians. As such, their DOAs, 
$\phi_{\ell}$ and $\phi_{k}$ are independently, and uniformly drawn from $\left[\frac{-\pi}{3},\frac{\pi}{3}\right]$. From 
\eqref{effectiveinterferenceapproximation}, the probability density function of $\Theta_{\ell,k}$ with $M$ antennas at the lens array 
can be written as 
\begin{equation}
f_{\Theta_{\ell,k}}^{\left(M\right)}\left(z\right)\approx
\int\nolimits_{y_{L}}^{y_{U}}\hspace{-1pt}\frac{1}{\tilde{D}}\left(\frac{3}{2\pi}\right)^{2}\hspace{-5pt}
\frac{1}{\sqrt{1-\left(\frac{z}{\tilde{D}}+y^{2}\right)}}\frac{1}{\sqrt{1-y^{2}}}\hspace{5pt}dy; \hspace{15pt}\left|z\right|\leq{}\sqrt{3}\tilde{D},
\end{equation}
where the bounds on the integral are defined given by $y_{L}=\max\left(-\frac{\sqrt{3}}{2},-\frac{\sqrt{3}}{2}-\frac{z}{\tilde{D}}\right)$, and $y_{U}=\min\left(\frac{\sqrt{3}}{2},\frac{\sqrt{3}}{2}-\frac{z}{\tilde{D}}\right)$. Then, following 
\eqref{effectiveinterfererprob}, the probability can be written as 
\begin{equation}
\label{effectiveinterfererprob2}
p_{k}\approx\int\nolimits_{-1}^{1}f_{\Theta_{\ell,k}}^{\left(M\right)}\left(z\right)\hspace{5pt}dz. 
\end{equation}
Applying the standard Taylor series expansion of \eqref{effectiveinterfererprob} and performing some straightforward simplifications yields 
\begin{align}
\nonumber
p_{k}&\approx\frac{9}{\tilde{D}\pi^{2}}\left[\tanh^{-1}\left(\frac{\sqrt{3}}{2}\right)+\mathcal{O}\left(\frac{1}{M^{2}}\right)\right]\\
\label{effectiveinterferenceprob3}
&\approx\frac{9}{\tilde{D}\pi^{2}}\left[\tanh^{-1}\left(\frac{\sqrt{3}}{2}\right)\right]. 
\end{align}
This yields the desired expression in \eqref{effectiveinterfererprobability}, and concludes the proof. \QED


\end{document}